\documentclass[10pt, paper=a4, UKenglish]{article}
\usepackage{graphicx}
\usepackage[frozencache,cachedir=.]{minted}

\def\Title#1{\begin{center} {\Large #1 } \end{center}}
\def\Author#1{\begin{center}{ \sc #1} \end{center}}

\newcommand\pubblock{\rightline{\begin{tabular}{l} Proceedings of the CTD 2022\\ \pubnumber\\
         \pubdate  \end{tabular}}}

\newenvironment{Abstract}{\begin{quotation} \begin{center} 
             \large ABSTRACT \end{center}\bigskip 
      \begin{center}\begin{large}}{\end{large}\end{center} \end{quotation}}

\newenvironment{Presented}{\begin{quotation} \begin{center} 
             PRESENTED AT\end{center}\bigskip 
      \begin{center}\begin{large}}{\end{large}\end{center} \end{quotation}}

\def\Acknowledgements{\bigskip  \bigskip \begin{center} \begin{large}
      \bf ACKNOWLEDGEMENTS \end{large}\end{center}}

\def\beq{\begin{equation}}
\def\eeq#1{\label{#1}\end{equation}}
\def\eeqn{\end{equation}}

\def\beqa{\begin{eqnarray}}
\def\eeqa#1{\label{#1}\end{eqnarray}}
\def\eeqan{\end{eqnarray}}

\let\bar=\overbar

\def\Dslash{\not{\hbox{\kern-4pt $D$}}}
\def\dslash{\not{\hbox{\kern-2pt $\del$}}}

\def\msb{{\bar{\ssstyle M \kern -1pt S}}}

\usepackage{color}
\usepackage{lineno}
\usepackage{subfig}
\usepackage{hyperref}
\usepackage{authblk}

\newcommand\pubnumber{PROC-CTD2022-29 \\ LHCb-PROC-2022-012}

\newcommand\pubdate{\today}

\def\collaboration{\begin{center}On behalf of the LHCb Collaboration
\end{center}}
\def\affil#1{\begin{center} { \it #1} \end{center}}

\newcommand{\conference}{Connecting the Dots Workshop (CTD 2022)\\
May 31 - June 2, 2022}

\usepackage{fancyhdr}
\definecolor{mygrey}{RGB}{105,105,105}
\begin{document}


\large
\begin{titlepage}
\pubblock

\vfill
\Title{Fast and flexible data structures for the LHCb Run 3 software trigger}
\vfill

\Author{Arthur Hennequin \affil{Massachusetts Institute of Technology, USA}}
\Author{Michel De Cian \affil{École polytechnique fédérale de Lausanne, Switzerland}}
\Author{Sevda Esen \affil{University of Zurich, Switzerland}}

\collaboration
\vfill

\begin{Abstract}
Starting in 2022, the upgraded LHCb detector will collect data with a pure software trigger. In its first stage, reducing the rate from 30MHz to about 1MHz, GPUs are used to reconstruct and trigger on B and D meson topologies and high-$p_T$ objects in the event. In its second stage, a CPU farm is used to reconstruct the full event and perform candidate selections, which are persisted for offline use with an output rate of about 10~GB/s. Fast data processing, flexible and custom-designed data structures tailored for SIMD architectures and efficient storage of the intermediate data at various steps of the processing pipeline onto persistent media, e.g. tapes, is essential to guarantee the full physics program of LHCb. In these proceedings, we will present the event model and data persistency developments for the trigger of LHCb in Run 3. Particular emphasize will be given to the novel software-design aspects with respect to the Run 1+2 data taking, the performance improvements which can be achieved and the experience of restructuring a major part of the reconstruction software in a large HEP experiment.
\end{Abstract}

\vfill

\begin{Presented}
\conference
\end{Presented}
\vfill
\end{titlepage}
\def\thefootnote{\fnsymbol{footnote}}
\setcounter{footnote}{0}
%

\normalsize 


\section{The LHCb upgrade}
 In the last years, the LHCb detector\cite{Alves:2008zz} has been upgraded to run at a five times higher instantaneous luminosity ($\mathcal{L}$) than during Run 1+2 of the LHC, corresponding to $\mathcal{L} = 2\cdot 10^{33}\, \textrm{cm}^{-2} \textrm{s}^{-1}$. All tracking detectors and most of the readout electronics of the subdetectors have been replaced\cite{Collaboration:1647400,Collaboration:1624070,LHCb-TDR-014}. The change in hardware is also reflected by a change in the trigger strategy: for Run 3, starting in 2022, LHCb will use a pure software trigger, processing events at the 30 MHz non-empty bunch-crossing rate\cite{CERN-LHCC-2014-016}. The first stage of the software trigger, HLT1, uses a set of GPU cards to perform a partial event reconstruction\cite{LHCbCollaboration:2717938,Boettcher:2823780, Scarabotto:2823783}. HLT2, the second stage of the software trigger, employs a farm of CPU servers to fully reconstruct the event with offline quality and perform an event selection, given an input rate of about 1 MHz and an output data rate of about 10~GB/s\cite{CERN-LHCC-2018-014,Guenther:2819858}. 

In order to cope with the high data rate and the throughput requirements, the HLT2 event model was rewritten to use modern software paradigms such as SIMD (single instruction, multiple data) instructions. Its different building blocks and the performance will be explained in the following sections.

\section{Introduction to the LHCb event model}
\label{intro}


The LHCb event model consists of all classes, implemented in \texttt{C++}, that represent the data flow from the detector raw banks to the charged and neutral particles used for data analysis. It is used to pass information between the algorithms in the reconstruction chain and to consistently write and read information from and to files.  
In Run 1+2 of the LHC, the LHCb event model used so-called ``keyed containers" where every object in a container is identified by a key. These containers were implemented as array of structures (AOS) leading to slow data access in parallel-processing environments. Additionally, the keyed containers held pointers to objects they contained, making memory allocation and de-allocation slow. 
 
\begin{figure}[!htb]
  \centering
 	\includegraphics[trim=0.cm 9.5cm 0cm 0cm,clip, width=1.\linewidth]{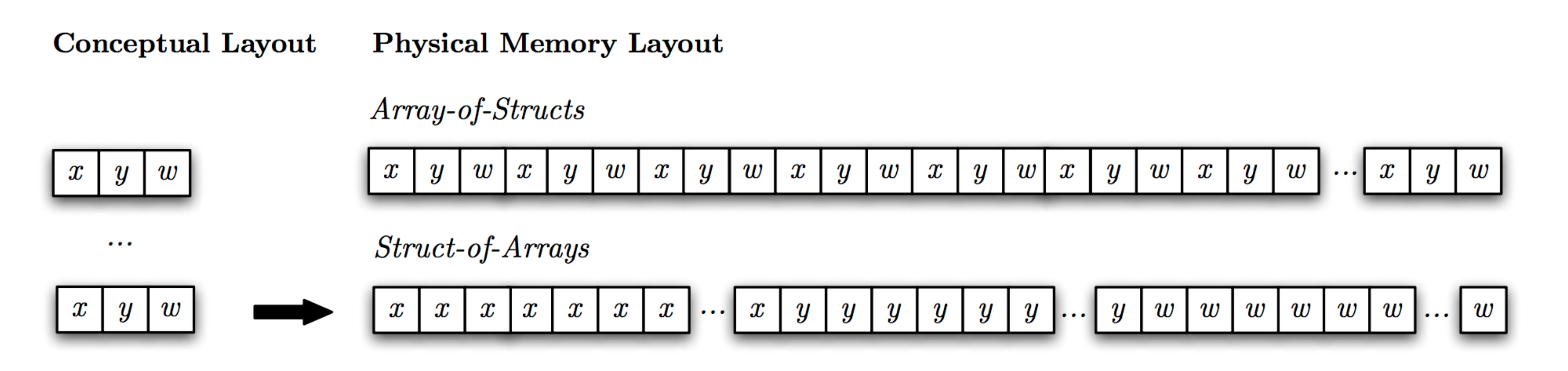}
  \caption{A comparison of AOS and SOA layouts. Taken from Ref.~\cite{Mniszewski_2021}.}
  \label{fig:aos-soa}
\end{figure}

For Run 3 of the LHC, the  pure software trigger of LHCb required a redesign of the event model to reach the desired throughput. The new model stores the data in an Struct-of-Arrays (SOA) layout to be able to take advantage of SIMD (single input, multiple data) instructions on CPUs. The memory layout of AOS and SOA structures can be seen in Fig.~\ref{fig:aos-soa}.
      
While developing the new event model, several key points for the event reconstruction, but also for the analysis of particle decays, have been taken into account. They include  having flexible data structures that can be grown and shrunk at run time using dynamic memory allocation, but also the possibility of traversing decay trees for the analysis of multi-staged particle decays. In order to reach a high computational speed, the model needs to allow easy vectorisation\cite{lemaitre:hal-01361204,lemaitre:hal-01550129,lemaitre:hal-01760260,Hennequin_2020}. At the same time, the new model needs to be compatible with the old event model also during the development phase to not break the workflow of the full reconstruction sequence and for quality assurance.

\section{SOA collections}
\label{soa-collections}

An SOA collection is a dynamically-resizeable collection of arrays in an SOA layout. Each array or field is represented by a tag which carries all the information about the field: its type, its packed representation for offline storage, etc... For example, a simple track\footnote{A track represents the trajectory of a charged particle.} collection can be created with:

\begin{minted}[mathescape]{c++}
// Define tags:
struct Momentum : float_field {};
struct Index : int_field {};
struct LHCbID : lhcbid_field {};
struct Hits : vector_field<struct_field<Index, LHCbID>> {};

// Define collection:
struct Tracks : 
     SOACollection <Tracks, Momentum, Hits>{};
\end{minted}

with \texttt{Momentum} the absolute momentum of the track, \texttt{LHCbID} a unique identifier for a charged cluster on a track, \texttt{Index} the index of the LHCbID on the track and \texttt{Hits} a class representing the collection of charged clusters on the track.
The design goal of SOA collections is to provide a user friendly structure, replacing an AOS structure such as \texttt{std::vector<Track>}, that allows for efficient vectorization. 

Access to the individual tags is provided via proxies, where the specific SIMD\footnote{SIMD is used in the following for all architectures which provide a vector width larger than 1.} or scalar backends can be chosen at compile time, with an automated detection of the largest vector width available on the specific architecture. A proxy therefore represents a chunk of N objects in the collection where one object, e.g. a track, is a slice through the collection. 

Elements to the collection can be easily added to the end, similarly to a \texttt{std::vector}, with the possibility of masking some elements, i.e. not actually adding them. This allows for selecting some objects while discarding others in parallel, e.g. when applying track quality or momentum requirements.

\begin{minted}[mathescape]{c++}
// Push N elements to the end of tracks, masking some
// Set the momentum of the track
auto proxy = tracks.emplace_back <simd>(mask); 
proxy.field<Momentum>().set(momentum);

// Iterate over tracks N elements at a time
for (const auto& proxy : tracks.simd()) 
    auto momentum = proxy.get<Momentum>();
\end{minted}
 The same operations in scalar: 
\begin{minted}[mathescape]{c++}
// Push 1 element to the end of tracks, possibly masking it
// Set the momentum of the track
auto proxy = tracks.emplace_back <scalar>(mask); 
proxy.field<Momentum>().set(momentum);

// Iterate over tracks one at a time
for (const auto& proxy : tracks.scalar()) 
    auto momentum = proxy.get<Momentum>();
\end{minted}

\section{Connecting SOA Collections}
\subsection{Zipping}
\label{sec:zipping-relations}
\begin{figure}[!htb]
  \centering
 	\includegraphics[trim=0.cm 2cm 0cm 0cm,clip, width=.9\linewidth]{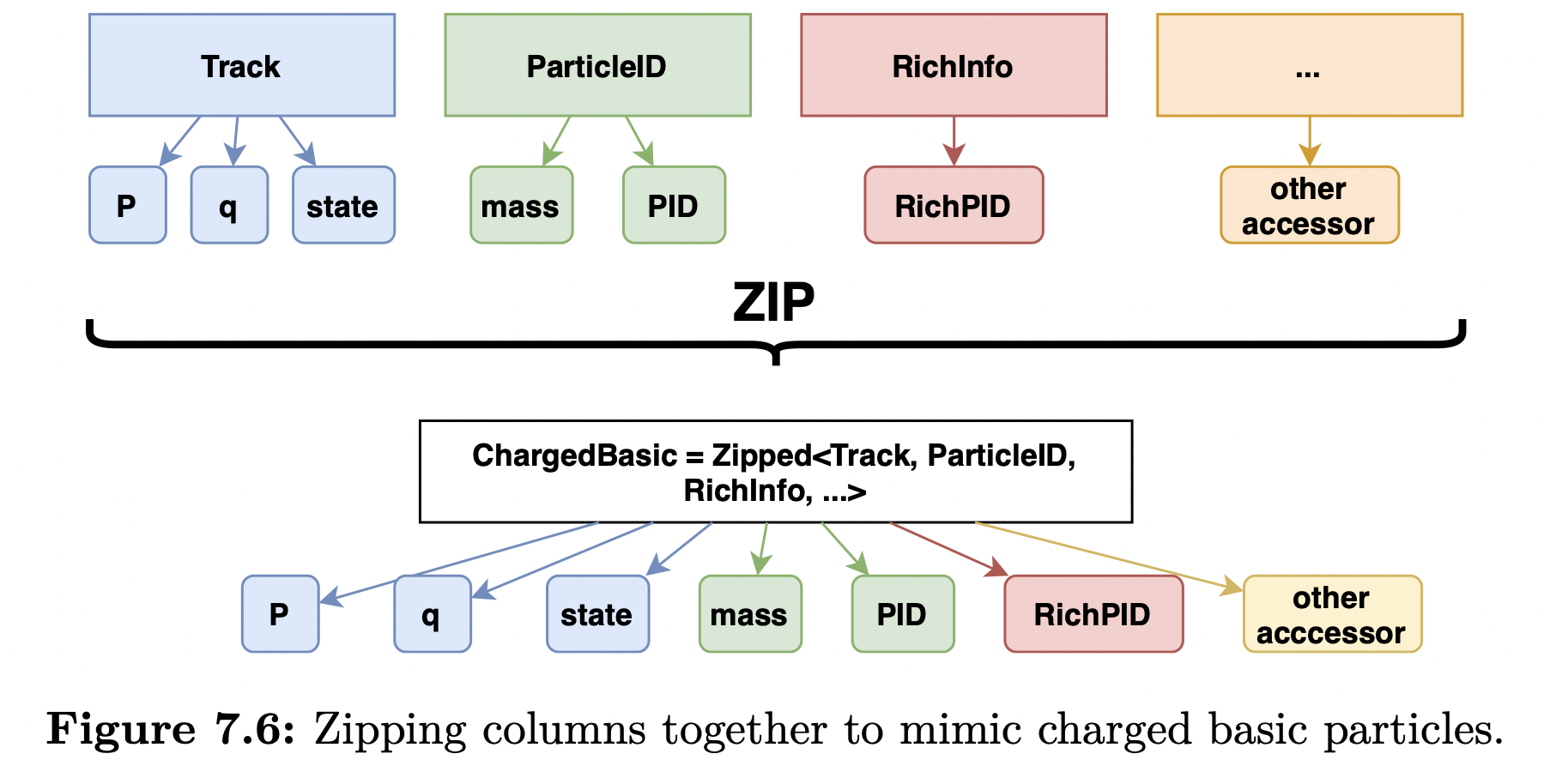}
  \caption{An example zip combining track, particle ID and RICH PID to a charged particle. Taken from Ref.~\cite{Nolte:2765896}.}
  \label{fig:zipping}
\end{figure}

In the transient event store (TES)\cite{Barrand:467678}, which is used to pass objects from one algorithm to the next, data objects need to be constant to allow safe memory allocation for multi-threading. However during the event reconstruction, more information can become available for some objects which are already in the TES. For example, after tracks are reconstructed, particle identification (PID) algorithms are executed, providing additional information for these tracks. Instead of making a copy of the objects in the TES, two methods can be used to connect the new information to the original object. The first is using `zipping', which is similar to python \texttt{zip()}.  A zip is a set of SOA collections of the same size that can be iterated as one and carries the information on how to iterate and access the collection, i.e. the actual SIMD backend and the proxy behaviour. An example of a possible zip between tracks and PID information can be seen in Fig.~\ref{fig:zipping}. Zips only keep pointers to existing containers and do not own any memory. An (example) zip with tracks and PIDs can be created with and iterated over with:
		
\begin{minted}[mathescape]{c++}
auto zipped = make_zip<simd>(tracks, PIDs);
for (const auto& zipproxy : zipped).simd() {
    auto momentum = zipproxy.get<Momentum>(); // from tracks 
    auto pid = zipproxy.get<pid>(); // from PIDs
}
\end{minted}

The fact that the code for looping over an SOACollection or a zip of SOACollections is identical leads to increased code flexibility.

\subsection{Relation tables}
Zipping only works if both SOA collections have the same size and there exists a one-to-one correspondence between the individual entries in the SOA collections. However, there are situations where this is not the case, i.e. two tracks could both point to the same calorimeter cluster.

The second method to add information to an existing object are therefore `relations'. Relations connect elements in a collection to something else, which can be another collection. An additional weight information can be added to each relation. SOA Relations are SOA Collections representing relations between two SOA Collections. For example a relation can be used between particles and their primary vertices with:

\begin{minted}[mathescape]{c++}
struct TracksPVsRelWithWeight: 
     RelationTable2D<Tracks, PVs, Weight>{};
TracksPVsRelWithWeight table {tracks,  pvs};
auto proxy = table.emplace_back<simd>();
proxy.set(tracks.indices(), pvs.indices(), weight);
\end{minted}

\section{SIMD Wrappers}
\label{simd-wrappers}	
The efficient use of SIMD instructions relies on using ``intrinsics" for vector operations, which depend on the architecture and instruction set used (x86, ARM; SSE, AVX). In order to allow for a consistent use of vector operations, an easy switching between the backends and a more familiar look-and-feel similar to the scalar instructions formerly used in the LHCb code, wrapper classes for commonly used intrinsics, called \texttt{SIMDWrapper} were introduced at LHCb\cite{hennequin:tel-03640612}. The instruction set is fixed at compile time, by selecting an architecture using compiler flags and target, to allow the compiler to do more optimizations. Given that changing the architecture during runtime is unlikely, this limitation does not have a negative impact for the LHCb software. The wrapper is fully integrated into the LHCb software and templated when possible to have only one implementation for all backends. Also common math functions and matrix operations are defined for all architectures to allow easy switching from one to another. An example for the function to find the minimum is given below:

\begin{minted}[mathescape]{c++}
// scalar
scalar::float_v min( scalar::float_v lhs, scalar::float_v rhs ) { 
    return std::min( lhs.data, rhs.data ); 
}
// neon
neon::float_v min( neon::float_v lhs, neon::float_v rhs ) { 
    return vminq_f32( lhs, rhs ); 
}
// avx
avx::float_v min( avx::float_v lhs, avx::float_v rhs ) { 
    return _mm256_min_ps( lhs, rhs ); 
}
\end{minted}
with \texttt{scalar::float\_v} a float with vector width one, \texttt{neon::float\_v} a float on the ARM architecture, \texttt{vminq\_f32} the function to find the minimum between two ARM float numbers,  \texttt{avx::float\_v} a float in the AVX instruction set and \texttt{\_mm256\_min\_ps} the function to find the minimum between two AVX float numbers.

\section{Throughput Oriented (ThOr) selections}
\label{sec:thor}
In Run 2, the event reconstruction at LHCb was about 70\% and the selections were about 30\% of the time spent in HLT2. This is expected to be similar in Run 3. Currently, more than 1000 exclusive HLT2 lines are being tested, each performing selections (cuts, vertex fitting, combinations etc...) on basic particles. In order to benefit from the speed improvement provided by SIMD instructions and the usage of SOA collections also in selections, a new framework was developed simultaneously for the old and the new SOA-based event models. 

In order to select interesting decays in trigger lines, functors (i.e. function objects) are used. The so-called Throughput Oriented (ThOr) functors, are designed to be agnostic about the input and output type to be flexible on what they operate on. 
A significant gain in speed is achieved when using SIMD instructions on SOA containers compared to the old implementation as seen in Table~\ref{tab:ThOr}. Additional speed is gained by using a functor cache instead of Just-In-Time compilation: This means that functors, which are defined in \texttt{python}, are compiled into a cache during the build process to be then used directly in the application without further interpretation. To simplify user experience, functors are  templated and are using SIMDwrappers, so the code is the same for every architecture and no specialization is needed at the functor level.

\begin{table}[!htb]
  \centering
 	\includegraphics[trim=0.cm 0cm 0cm 0cm,clip, width=.6\linewidth]{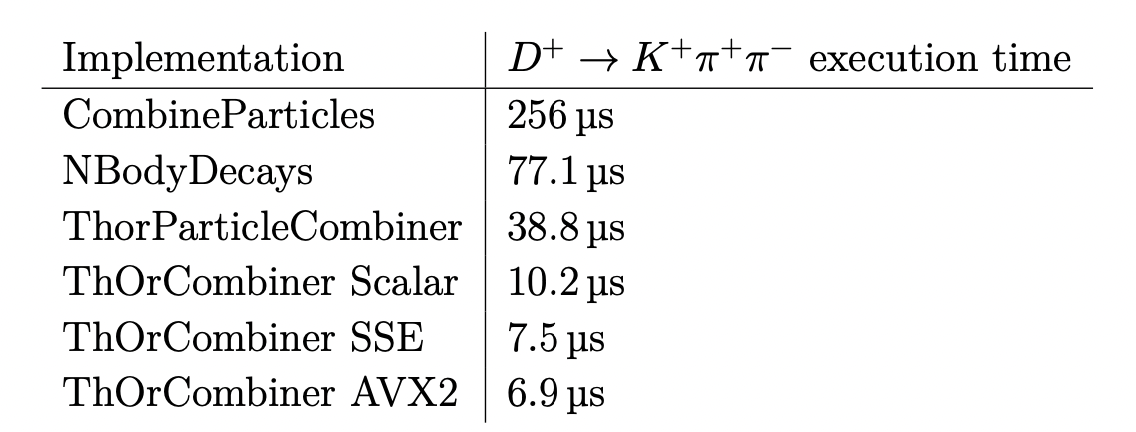}
  \caption{A benchmark comparing the timing of the reconstruction of a $D^{+} \to K^{+}\pi^{+}\pi^{-}$ particle decay, using different algorithms to perform the particle combination. CombineParticles and NBodyDecays use the legacy framework from Run 1+2; ThOrParticleCombiner uses functors, but still the old data structures; ThOrCombiner uses SOA structures with different vector widths. Taken from Ref.~\cite{Nolte:2765896}.}
  \label{tab:ThOr}
\end{table}

\section{Persistency}
\label{sec:persistency}

 The final step of the event reconstruction and the selection of candidates is the persistency of the data for future use. In the AOS event model, this is done in two steps: filtering what needs to be persisted and creating persistent representations i.e., conveying the data to more basic data structures. The SOA collections are  already mostly in a format that is ready to be persisted so the second step is simpler. While creating the collection, each tag can be customized for how and if it is persisted and versioning can be introduced. For the example below, one field is defined to be packed as float, one field is not to be persisted at all, and one field is to be persisted only for the newest versions of the collection. 
\begin{minted}[mathescape]{c++}
// Define tags :
struct Momentum : float_field {
using packer_t = SOAPackFloatAs <short,
                                std::ratio<1, 100 > >;
};
struct Unwanted : int_field {
   using packer_t = SOADontPack ;
};
struct OopsIForgotThisField : int_field {
using packer_t =
    SOAPackIfVersionNewerOrEqual<1, SOAPackNumeric <int>>;
};
// Define collection :
struct Tracks : SOACollection<Tracks,Momentum, 
                            Unwanted, OopsIForgotThisField> {};
\end{minted}

\section{Throughput improvements in the HLT1 prototype sequence}
A prototype for the HLT1 trigger was implemented on CPU in parallel to the GPU prototype\cite{LHCb:2021kxm}. It featured most of the improvements explained in these proceedings, most importantly the SOA Collection and a widespread use of SIMD instructions. The HLT2 trigger uses same event model as the HLT1 and benefits from the same improvements mentioned in these proceedings. 

In order to test the impact of using SIMD instructions compared to scalar instructions, the sequence was once run with the AVX2 instruction set, and once with scalar instructions, while keeping the event model the same. The throughput was about twice when using vector instructions compared to scalar instructions.

\begin{figure}[!htb]
  \centering
 	\includegraphics[trim=0.cm 0cm 0cm 0cm,clip, width=.8\linewidth]{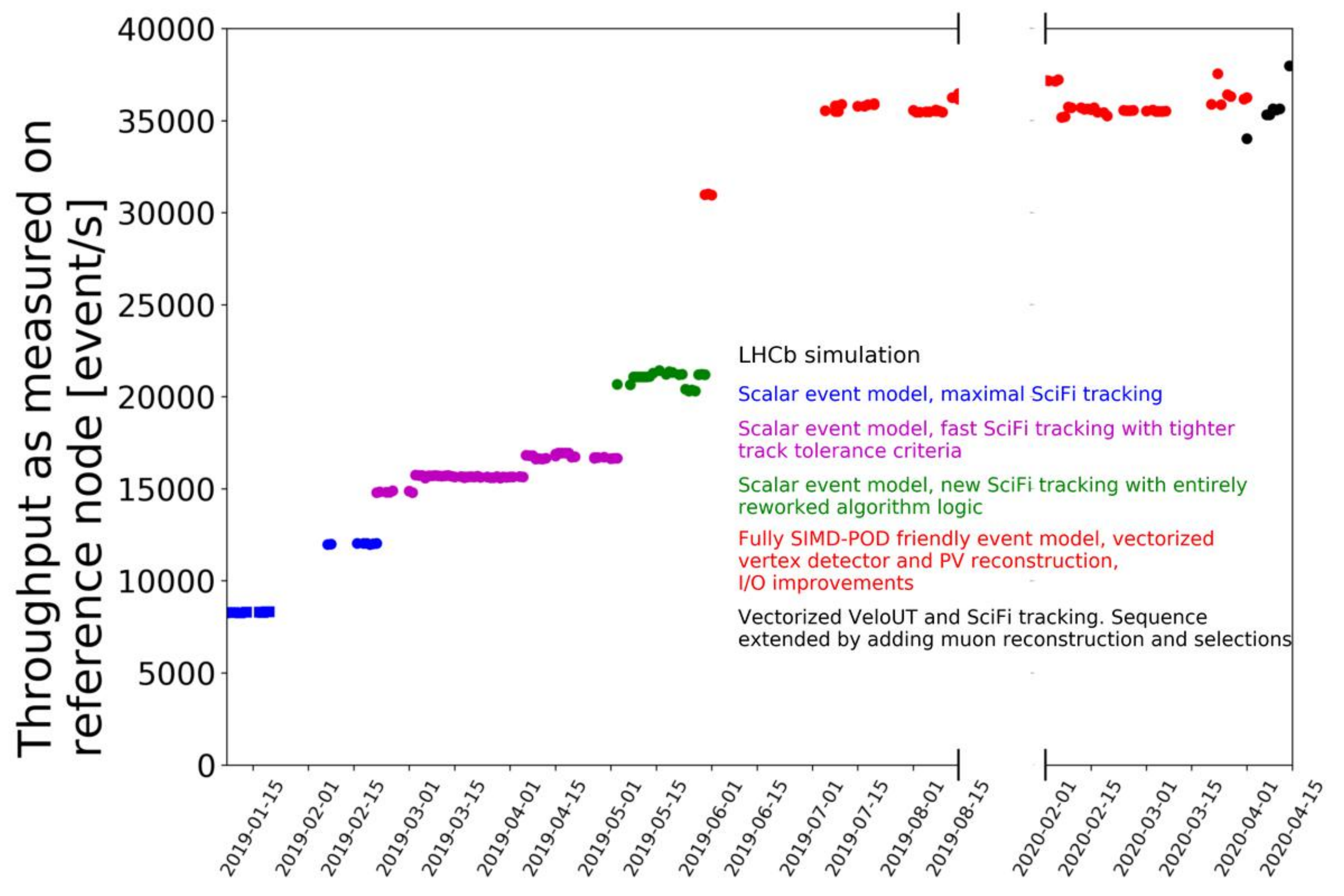}
  \caption{Evolution of the upgrade LHCb HLT1 throughput of the CPU prototype between December 2018 and April 2020. The period between August 2019 and February 2020 has been cut out as there were little changes to the throughput. More details about the indicated optimizations can be found in Ref.~\cite{LHCB-FIGURE-2020-007}.}
  \label{fig:HLT1CPU}
\end{figure}

A historical overview over the improvements achieved using the new event model, using SIMD instructions and having improved reconstruction algorithms can be seen in Fig.~\ref{fig:HLT1CPU}. It shows that thanks to a concentrated effort on all three points, the throughput could be improved by about a factor 4, without compromising the reconstruction of physics quantities.

\section{Conclusions}
For Run 3 of the LHC, the LHCb collaboration implemented a new event model for the second stage of the software trigger, HLT2, using an SOA layout, native usage of SIMD instructions and more flexibility. This results in an increased  throughput and allows to run more than 1000 trigger lines with a full offline-quality reconstruction, without the need for any post-processing. This event model therefore is well suited for the coming decade of data taking of the LHCb experiment.

\Acknowledgements

The authors would like to thank the LHCb computing, simulation and RTA teams for their support and for producing the simulated LHCb samples used for these proceedings. We also would like to thank the organisers of the CTD conference for the interesting workshop, nice location and the copious amount of delicious cookies.

M. De Cian acknowledges support from the Swiss National Science Foundation grant ``Probing right-handed currents in quark flavour physics", PZ00P2\_174016.
S. Esen acknowledges funding from the European Union’s Horizon 2020 research and innovation programme under the Marie Skłodowska-Curie grant agreement No. 101027131.


\bibliographystyle{unsrt}
\bibliography{refs} 


\end{document}